\documentclass[10pt,twocolumn,letterpaper]{article}

\usepackage{iccv}
\usepackage{times}
\usepackage{epsfig}
\usepackage{graphicx}
\usepackage{amsmath}
\usepackage{amssymb}

\usepackage[pagebackref=true,breaklinks=true,letterpaper=true,colorlinks,bookmarks=false]{hyperref}

\iccvfinalcopy 

\def\iccvPaperID{13} 

\ificcvfinal\fi

\usepackage[symbol]{footmisc}
\begin{document}

\title{Improving Tuberculosis (TB) Prediction using Synthetically Generated Computed Tomography (CT) Images}



\author{Ashia Lewis\\
The University of Alabama
\and
Evanjelin Mahmoodi\\
University of California, Santa Cruz\\
\and
Yuyue Zhou\\
New York University\\
\and
Megan Coffee\\
NYU Grossman School of Medicine\\
\and
Elena Sizikova \\
New York University\\
}

\maketitle
\ificcvfinal\thispagestyle{empty}\fi

\begin{abstract}
The evaluation of infectious disease processes on radiologic images is an important and challenging task in medical image analysis. Pulmonary infections can often be best imaged and evaluated through computed tomography (CT) scans, which are often not available in low-resource environments and difficult to obtain for critically ill patients. On the other hand, X-ray, a different type of imaging procedure, is inexpensive, often available at the bedside and more widely available, but offers a simpler, two dimensional image. We show that by relying on a model that learns to generate CT images from X-rays synthetically, we can improve the automatic disease classification accuracy and provide clinicians with a different look at the pulmonary disease process. Specifically, we investigate Tuberculosis (TB), a deadly bacterial infectious disease that predominantly affects the lungs, but also other organ systems. We show that relying on synthetically generated CT improves TB identification by 7.50\% and distinguishes TB properties up to 12.16\% better than the X-ray baseline.

\end{abstract}

\section{Introduction}

Diagnosing and evaluating infections on medical images is an essential role of medical imaging. Computed tomography (CT) scans provide substantial insight into diagnosis and follow-up of patients already on treatment, uncovering details on the patient's particular case. However, CT scanners are expensive devices that are often not available in low-resource, smaller or rural communities~\cite{world2011baseline}, not always feasible for critically ill patients, and present a large dose of radiation. Even if the initial diagnosis was made with CT, follow up exams often cannot happen due to accessibility, costs, and concerns over radiation dose. This is an important radiology challenge which precludes millions of people worldwide from receiving an advanced medical evaluation~\cite{mollura2014radiology,ngoya2016defining}. For deadly infectious diseases such as tuberculosis (TB) that particularly affect those in lower resource settings, clinicians are often left without the same radiologic insights that clinicians in resource rich settings may have for patient care. CT scans, when available, may not have as many imaging slices or may have delays because of needed repairs or the need for trained radiology technicians and radiologists to be involved. Improved access to radiology with synthetic scans could also increase clinician engagement. Tuberculosis is a major public health concern, and year after year has been the largest infectious disease cause of death worldwide. Timely diagnosis and treatment, with careful follow-up, is not only imperative in saving patient lives, but also critical for preventing disease spread across the larger community.

\begin{figure*}[ht]
\centering
  \includegraphics[width=1.0\linewidth]{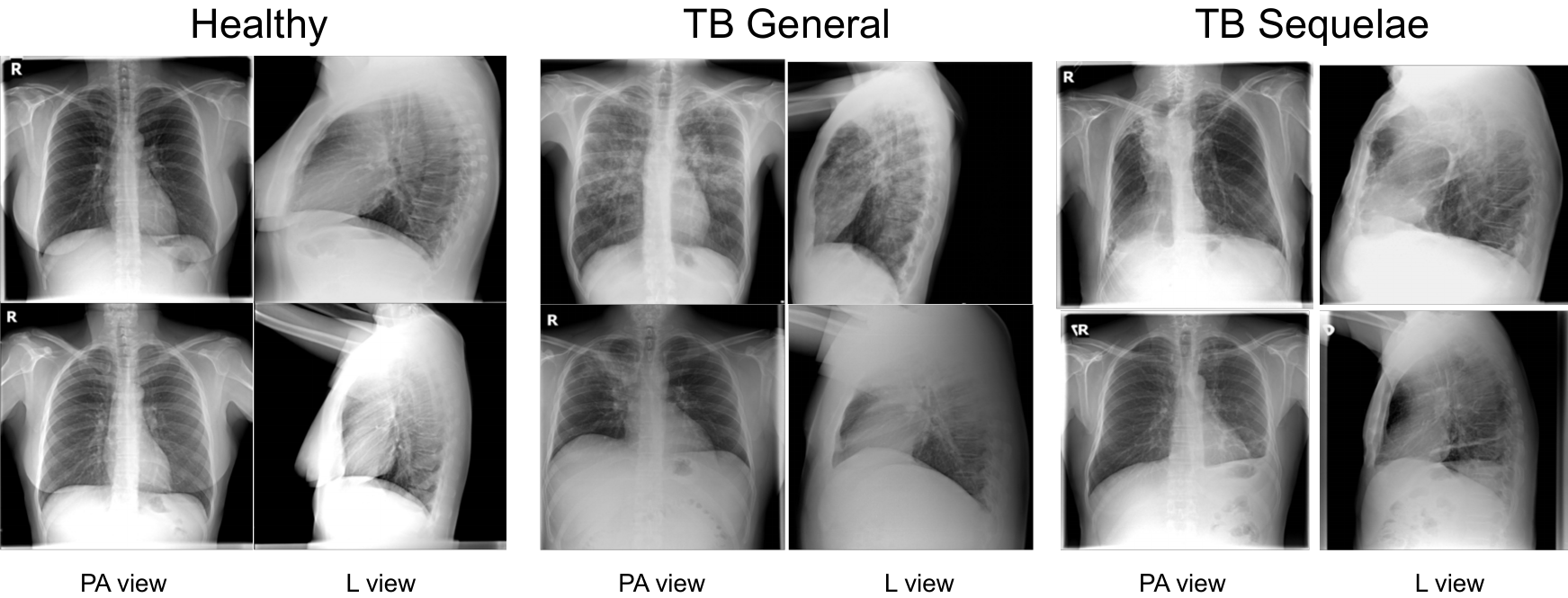}
  \caption{Examples of X-ray (posterior-anterior (PA) and lateral (L) ) of healthy and diseased patients. In comparison to healthy patients (first column), lungs of TB patients (second column) may have many different findings. Here is shown one specific finding: infiltrates (areas gray to white) in the upper lung fields which represent active infection. Other examples of active TB may look very different, such as speckled pattern throughout the lung (miliary TB named as appears as if millet thrown across image) and effusions (fluid appearing white on imaging) at the lung base(s). TB sequelae~\cite{machida2005state} (Sequela of tuberculosis) refers to scarring in the absence of infection after healing from TB. This can have many different findings as well. There may be a decrease in lung volumes, airways may narrow, and there may be calcification (deposition of calcium) or cavitation (loss of or holes in lung parenchyma), as described in Figure 2.
  }
  \label{fig:examples}
\end{figure*}

In this work, we leverage the power of computational modelling to estimate CT scans from chest radiographs (X-rays), a different type of imaging procedure. Each X-ray provides a single image from one angle, creating a 2D image. CT scans take the same imaging technique with multiple images, generally using a rotating X-ray tube, creating what is essentially a 3-D image with multiple image slices available. Image quality can vary in different settings and many X-rays, performed without quality equipment or trained technicians, may be blurry and may be under or over-exposed. X-rays are more widely available, cost-effective, may even be done at the bedside for some patients, and require less radiation. We use the X-ray and CT images to learn an automatic disease classification network, focusing on healthy and tuberculosis (TB) patients, as well as different TB properties (see Figure~\ref{fig:examples}), and show that synthetically generated CT scans improve identification of TB by 7.50\% and classification of TB properties by 12.16\%. These synthetic CT scans also give tangible imaging for clinicians who may not have access to CT scans to use in order to diagnose and follow patients clinically.

We generate CT scans from input X-ray images using an existing computational model, X2CT-GAN~\cite{ying2019x2ct}. This model is a neural network trained to predict CT from chest radiographs (X-ray). By relying on digitally reconstructed radiograph (DRR) technology~\cite{milickovic2000ct}, this technique obtains synthetic training X-rays in posterior-anterior (PA) and optionally lateral views (see Section~\ref{sec:approach} for discussion) and learns a generative adversarial network (GAN) to predict the CT image from input X-rays. The model is further refined to accept the distribution of real X-rays by style transfer using CycleGAN~\cite{zhu2017unpaired}. Please see \cite{ying2019x2ct} for additional details. In this work, we study how CT scans generated using this model can improve the identification and classification of tuberculosis (TB). In particular, we learn a fully automatic tuberculosis classification model and study how different types of inputs affect disease identification and classification performance. Our contributions can be summarized as follows:
\begin{itemize}
\item We show that it is possible to significantly improve automatic TB disease identification and classify its variants by generating synthetic CT scans from X-ray images before the analysis phase.  

\item We conduct an extensive experimental evaluation of our approach on disease identification and property classification, and present compelling quantitative and qualitative results.
\end{itemize}

\begin{figure*}[ht]
\vspace{0.5cm}
\centering
  \includegraphics[width=1.0\linewidth]{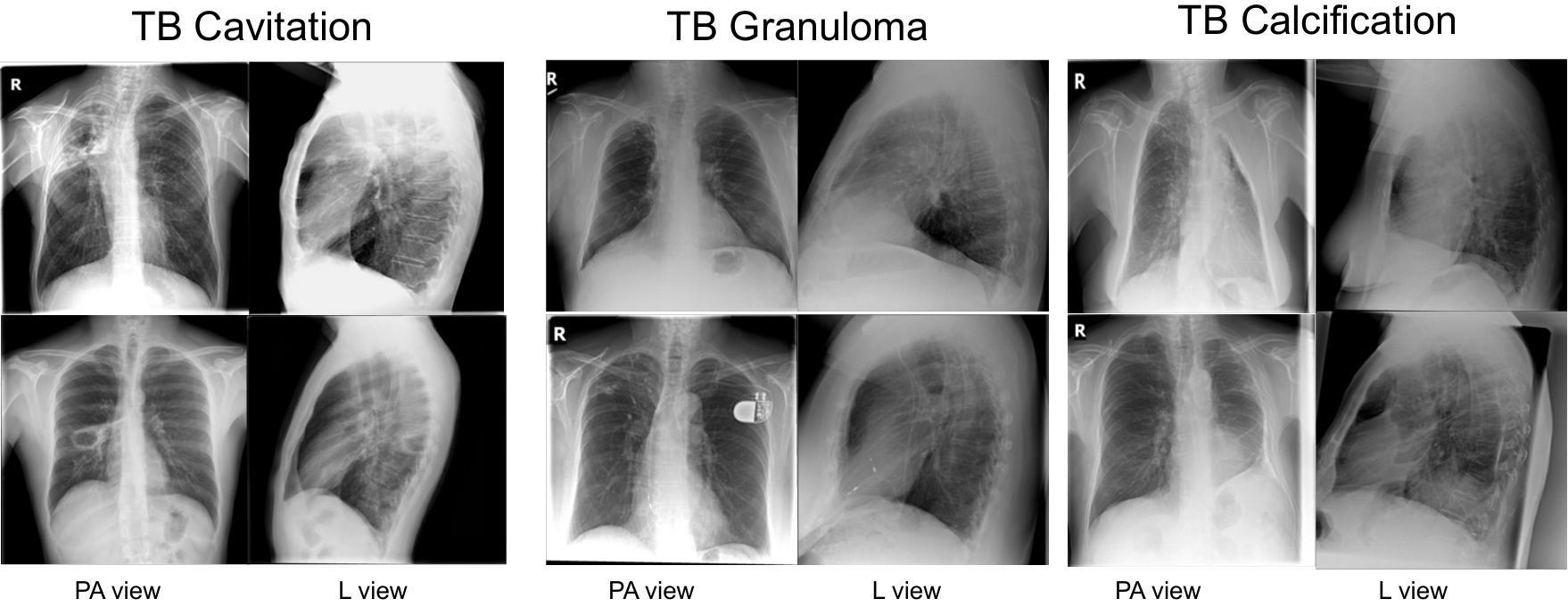}

  \caption{Examples of different types of tuberculosis (TB) categories considered: cavitation, granuloma, calcification.  Cavitations refer to cavities (holes) in the lung that can develop as a result of tuberculosis infection and are identified as dark, usually round areas on an X-ray. Granulomas refer to an inflammatory response and can often be identified by a nodule, generally white because of calcification. Calcifications refer to any area where TB has led to the deposition of calcium, resulting in areas white like bone, and this can include small flecks, larger nodules (such as granulomas), and also the pleural or pericardial (heart) lining}
  \label{fig:examples_categorizes}
\end{figure*}

\section{Related Work}

\subsection*{Automatic  Disease Classification}
The success of deep learning-based models in computer vision tasks~\cite{russakovsky2015imagenet} encouraged development of many automatic disease classification networks. Such approaches have been used to classify COVID-19~\cite{wang2020covid}, tuberculosis~\cite{pasa2019efficient}, and other infectious diseases~\cite{sharma2020artificial,wang2017hospital} from X-ray images. CT images~\cite{pham2020comprehensive,kuruvilla2014lung} have also been widely used for automatic disease classification, in part because analysis of 3D imaging offers additional context~\cite{zou20173d}. 
In particular, the value of CT scans for tuberculosis (TB) classification using neural networks has been recently highlighted~\cite{zunair2020uniformizing,kazlouski2019imageclef}. Both CT and X-ray can be used for diagnosis and analysis of TB, but CT may offer additional visualization and insight~\cite{nachiappan2017pulmonary,burrill2007tuberculosis,alkabab2018performance}. 

On the other hand, multi-modal models, for instance, computed tomography (CT) and electronic health records (EHR)~\cite{huang2020multimodal} or magnetic resonance imaging (MRI) and positron emission tomography (PET)~\cite{liu2018multi}, have been shown to successfully fuse information across multiple modalities. In a similar fashion, recent studies~\cite{hashir2020Quantifying,zhu2021mvc} identified that the addition of different views improves chest X-ray prediction accuracy. Fusion of X-ray and CT in particular has been shown to help automatically diagnose infectious diseases such as the recent coronavirus (COVID-19)~\cite{maghdid2021diagnosing}. 

\subsection*{Synthetic Data in Medical Imaging}
Development of realistic generative models, such as generative adversarial networks (GAN)~\cite{goodfellow2014generative}, made possible the generation of very realistic synthetic medical data such as patient records~\cite{choi2017generating}, skin~\cite{ghorbani2020dermgan} and bone~\cite{gupta2019generative} lesions, mammograms~\cite{korkinof2018high} and others~\cite{kazeminia2020gans}. Synthetically generated medical data has been used to address limitations on real data availability such as class imbalance~\cite{li2010learning} or limited availability~\cite{frid2018gan}. For example, in combination with real data, synthetic images are effective in improving localization of important anatomical landmarks~\cite{teixeira2018generating}. In another study~\cite{salehinejad2018generalization}, simulated chest X-rays with pathologies helped improve chest pathology classification. Synthetically generated CT images have been shown to improve a diverse set of medical data analysis tasks such as classification~\cite{frid2018gan}, segmentation~\cite{sandfort2019data}, super-resolution~\cite{you2019ct}. In our work, we rely on a recent GAN-based approach~\cite{ying2019x2ct} that generates CT images from X-rays to improve. To the best of our knowledge, we are the first to use synthetically-generated CT images for the analysis of TB.

\section{Tuberculosis Background}
TB imaging can have many findings. In our analysis, imaging with TB was classified as having granulomas, other calcifications, and cavitations (see Figure~\ref{fig:examples_categorizes} for examples), though there are many other findings associated with TB diagnosis. Imaging identified as having TB was also classified as either active TB or TB sequelae. Note that patients may have multiple classifications.

The different specific findings include granulomas and other calcifications. Granulomas refer to an inflammatory response, which encapsulates bacteria responsible for causing TB but also potentially in response to other triggers. Formally, granulomas are identified by pathological analysis, but on radiology these may be identified when there is a nodule, often calcified and hence white, on TB imaging. Other calcifications, where tuberculosis infection has led to the deposition of calcium, are also identified. This can include calcification of the pleural space, which is found surrounding the outer lung. Cavitations refer to holes in the lung which can develop in tuberculosis infection and persist indefinitely after. These cavities appear as darker, often rounded, areas on imaging.

TB sequelae~\cite{machida2005state} (Sequelae of tuberculosis) refers to the state of various complications after healing from TB, and is characterized by scarring of the lungs in the absence of signs of active infection. These sequelae can include: cavitations and various calcifications, as well as apical fibrosis (scarring or stiffening of the upper lung tips), cicatrization atelectasis (scarring of lung reducing the expansion or size of the lungs), tracheobronchial stenosis (narrowing of airways), or bronchopulmonary fistula (a passageway between airways and the pleural space that surrounds the lungs). Patients, without extensive medical records may be treated and retreated for findings of TB sequelae that are misdiagnosed as active TB. Such patients may have continued difficulty breathing or other symptoms but this will not respond to treatment, which can be harmful especially as there is no benefit as there is no active TB infection. In our analysis, we study how automatic classification of TB findings and identification of TB can be obtained from X-rays and improved using synthetically generated CT images.

\section{Approach} \label{sec:approach}
We leverage a generative model that predicts CT scans from X-rays in order to evaluate its effectiveness in improving TB prediction accuracy. We divide our analysis into two tasks: \emph{disease identification}, where the goal is to distinguish healthy and TB patients, and \emph{disease classification}, where the goal is to classify different types of TB or to distinguish active from past TB (TB sequelae).

Specifically, let $x_i = (x^{PA}_i,x^{L}_i,d_i)$ be the X-ray images (in posteroanterior (PA) and lateral (L) views) of a given patient, where $d_i = (l_i,s_i,t_i)$ is the disease label. Binary variables $l_i,s_i,t_i$ are such that: $l_i$ denotes whether the patient has TB, $s_i$ denotes whether the patient has TB sequelae, $t_i$ denotes whether the type of TB (granuloma, cavitation, calcification) that the patient has. Let $X=\{x_1,x_2,\ldots x_n\}$ be a collection of data available. We use computational models $f_1,f_2$ to predict $c_{1,i} = f_1(x^{PA}_i)$ and $c_{2,i} = f_2(x^{PA}_i, x^{L}_i)$, the synthetic CT image corresponding to example $i$. 

We then train separate models $g$ such that: $\hat{q}_i = g(r_i)$, where $q_i=l_i,s_i$ or $t_i$, and $r_i$ is a subset of $\{x^{PA}_i,x^{L}_i,c_{1,i}, c_{2,i} \}$. The loss function used for training $g$ is the standard binary cross entropy (BCE) with logits loss:
\begin{equation*}
\mathcal{L} (q_i,\hat{q}_i) = - [ (\hat{q}_i) \cdot \log \sigma (q_i) + (1- \hat{q}_i) \cdot (1-\sigma (q_i)) ]
\end{equation*}
where $q_i$ is the ground truth label and $g(r_i)$ is the predicted label. Note that $g(r_i)$ and $q_i$ are represented as binary (multi-dimensional) variables and above loss is applied element-wise with mean reduction in the end. 

All functions $f_1,f_2$ and $g$ were implemented using neural networks and described below. 

\section{Experimental Details}
We now describe the dataset and implementation details of our approach. We tested the proposed methodology on a large public benchmark~\cite{padchest}. In practice, $f$ and $g$ are implemented as neural networks and are trained as described below.  

\subsection{Dataset}
To experimentally analyze the proposed approach, we used the PAthology Detection in Chest radiographs (PadChest)~\cite{padchest} dataset, consisting of 160,868 X-ray images and medical reports from 67,000 patients. The dataset was collected and interpreted by 18 radiologists at the Hospital Universitario de San Juan, Alicante (Spain) from January 2009 to December 2017. About a quarter of the reports were manually annotated by radiologists and the rest were labelled using a supervised recurrent neural network model (achieving a 0.93 Micro-F1 score on an independent test). The dataset had 50.3\% female patients, 49.7\% male patients. The patient's ages ranged from 0 to 105 years, with a mean age of 58.5. For each experiment, we filtered the original datasets into smaller sets with required annotations. Specifically, for the TB identification experiment (healthy vs TB), we used 274 training, 91 validation, and 91 testing images. For TB classification into either TB or TB sequelae, we used 291 97 testing images. For TB classification into granuloma, cavitation, and calcification, we used 453 training, 123 validation, and 181 testing images. 

The X-ray images in this dataset contain at least one PA view, L view, or both (mean number of images per patient is 2.37). For multi-view (PA + L) experiments, we selected a corresponding pair of views. For the single-view (PA or L) experiments, we randomly selected a single image per patient.

\section{Network Architecture and Training}
\subsection{CT Generation Network}
We use the X2CT-GAN model~\cite{ying2019x2ct} to generate both biplanar (B) and single-view (SV) synthetic CT scans for input X-ray images. We relied on the publicly available pre-trained model for all of our experiments, which was obtained from the paper code repository. This model was trained on the LIDC-IDRI dataset~\cite{armato2011lung} that consists of 1,018 chest CT images, augmented with synthetic data (please see \cite{ying2019x2ct} for additional details). Both the posteroanterior (PA) and lateral (L) radiographic views were used to generate the biplanar synthetic CT images, while the single-view CT scans were generated using only the PA view. 

\subsection{TB Analysis Network}
For X-ray experiments, we resized images to resolution 364 by 364 with random resized crop of 320 and flip, and normalized them. We used the HeMIS model~\cite{havaei2016hemis} with  DenseNet-PA architecture based on implementation in \cite{hashir2020Quantifying} to perform classification. We use Adam as optimizer, BCE as loss function, set learning rate as 0.0001 and batch size of 1. The model was trained until convergence. 

For CT and CT+X-ray experiments with two classes, we relied on the 3D convolutional architecture proposed by \cite{zunair2020uniformizing} with Adam optimizer, learning rate 0.0001 and  batch size 20. The network was trained until convergence with binary cross-entropy (BCE) loss for 2-class experiments. In this experiment, X-ray images were normalized and resized to 128 by 128 resolution, and Gaussian blur was applied to de-noise. The CT images were normalized and resized to 128 by 128, and two central slices were chosen. We experimentally found that using two center slices yielded better results than the entire CT volume (see Table~\ref{tab:identification_acc}), possibly because the non-center slices of the generated CT often appeared blurrier. For the joint X-ray and CT experiments, the X-ray and CT images were concatenated before being inserted into the network. During training, random rotation augmentations were used. For 3-class experiments, we used the architecture proposed by \cite{ghaderzadeh2021deep} with the same pre-processing and parameters as described above.

\section{Experimental Results}
\noindent In this section, we report quantitative and qualitative analyses of the proposed methodology.  

\subsection{Disease Identification Evaluation}
\noindent We start by analyzing the performance of TB identification (distinguishing healthy and TB patients) accuracy across different types of inputs. Results are reported in Table~\ref{tab:identification_acc}. We find that when trying to predict TB from X-rays only, using both PA and L views yields the same prediction accuracy. Therefore, in subsequent experiments, we used only the PA views to represent X-ray input to the network. 

We next look at the TB prediction accuracy from generated CT scans. We find that CT images generated from both PA and L views (i.e., the biplanar (B) images) offered 31.25\%  improvement over CT images generated only from a single (PA) view. In particular, the single view CT prediction accuracy was worse than the X-ray prediction accuracy, but the biplanar CT prediction accuracy was better than X-ray prediction accuracy by 5.00\%. This analysis also verifies the conclusion in \cite{ying2019x2ct} that X2CT-GAN generates more accurate CT scans from multi-view X-rays. An interesting observation is that the L view helps with CT generation, but not directly with classification. Notice also that the using a two-slice subset of the generated CT instead of the full volume offers an improvement of 5.00\%-11.11\%, likely because of the better quality of the generated images in the center.

When combined with X-rays, the combination of single view CT (generated using PA X-ray) and X-ray yielded the same accuracy as X-ray only, suggesting that addition of CT did not degrade performance of the X-ray based prediction when limited information was available from CT. On the other hand, the biplanar CT and X-ray generated the top accuracy of 0.86, which represents a 7.50\% improvement over the X-ray baselines. It appears that the CT generation network is constrained to generate salient geometric and disease-identifying properties that are not learned by the network that predicts disease directly from X-rays. 

\begin{table}[h]
\begin{center}
\begin{tabular}{|l|c|c|}
\hline
\textbf{Data}             & \multicolumn{2}{c|}{\textbf{Accuracy}} \\ \hline \hline
X-ray (PA)       & \multicolumn{2}{c|}{0.80}     \\ 
X-ray (PA) + (L) & \multicolumn{2}{c|}{0.80}     \\ \hline
                 & \textbf{Full CT}      & \textbf{2-Slice CT}           \\ \hline
CT (SV)          & 0.58      & 0.64              \\ 
CT (B)           & 0.80      & \underline{0.84}        \\ 
CT (SV) + X-ray  & 0.72      & 0.80              \\ 
CT (B) + X-ray   & 0.80      & \textbf{0.86}     \\ \hline
\end{tabular}


\end{center}
\caption{\emph{Evaluation of disease identification (classification into healthy vs TB)} accuracy. Testing accuracies are reported on different types of X-ray and generated CT images with notation: B - biplanar, SV- single view. Top results are achieved when PA X-ray images are combined with biplanar CT images and when 2-slice subset of the CT image is used. Best result is in bold, the second best is underlined.}
\label{tab:identification_acc}
\end{table}

\subsection{Disease Classification Evaluation}
\noindent Next, we evaluate the models on classifying TB properties (granuloma, cavitation, and calcification). We report the quantitative results of these experiments in Table~\ref{tab:classification_acc}. We find that the biplanar CT images improve the accuracy by 2.70\% over the X-ray baseline, while the combination of biplanar CT and X-rays gives the highest accuracy of 0.83, which is 12.16\% better than the X-ray baseline. 

\begin{table}[h]
\begin{center}
\begin{tabular}{|l|c|}
\hline
\textbf{Data} & \textbf{Accuracy} \\
\hline\hline
X-ray          & {0.74}          \\
CT             & \underline{0.76}           \\ 
CT + X-ray       & \textbf{0.83}           \\ \hline
\end{tabular}
\label{tab:classification_acc}
\end{center}

\caption{\emph{Evaluation of TB type classification (granuloma, cavitation, calcification) accuracy.} Top results are achieved when X-rays are combined with generated biplanar CT images. Best result is in bold, the second best is underlined.}
\end{table}

\subsection{Distinguishing TB and TB Sequelae}

\noindent Finally, in Table~\ref{tab:classification_acc_sequela}, we study the performance of the model in distinguishing TB and TB sequelae. Similar to the previous experiment, we found that the combination of biplanar CT images and X-rays generated the highest accuracy of 0.81, followed by the result of using only CT images. In this case, the combination of X-ray and CT demonstrated an improvement of 5.20\%, and the CT only demonstrated an improvement of 2.60\%, over X-ray only accuracies. These experiments demonstrate the clear improvement in accuracy as a result of using a pre-trained synthetic CT image generator in augmenting the X-ray inputs with a CT scan. 

\begin{table}[h]
\begin{center}
\begin{tabular}{|l|c|}
\hline
\textbf{Data} & \textbf{Accuracy} \\
\hline\hline
X-ray        & {0.77}          \\
CT           & \underline{0.79}           \\ 
CT + X-ray       & \textbf{0.81}           \\ \hline
\end{tabular}
\end{center}
\caption{ \emph{Evaluation of TB type classification (TB vs TB sequelae) accuracy.} Top results are achieved when X-rays are combined with biplanar CT images. Best result is in bold, the second best is underlined.}
\label{tab:classification_acc_sequela}
\end{table}

\begin{figure}[h]
    \centering
    \includegraphics[width=1.0\linewidth]{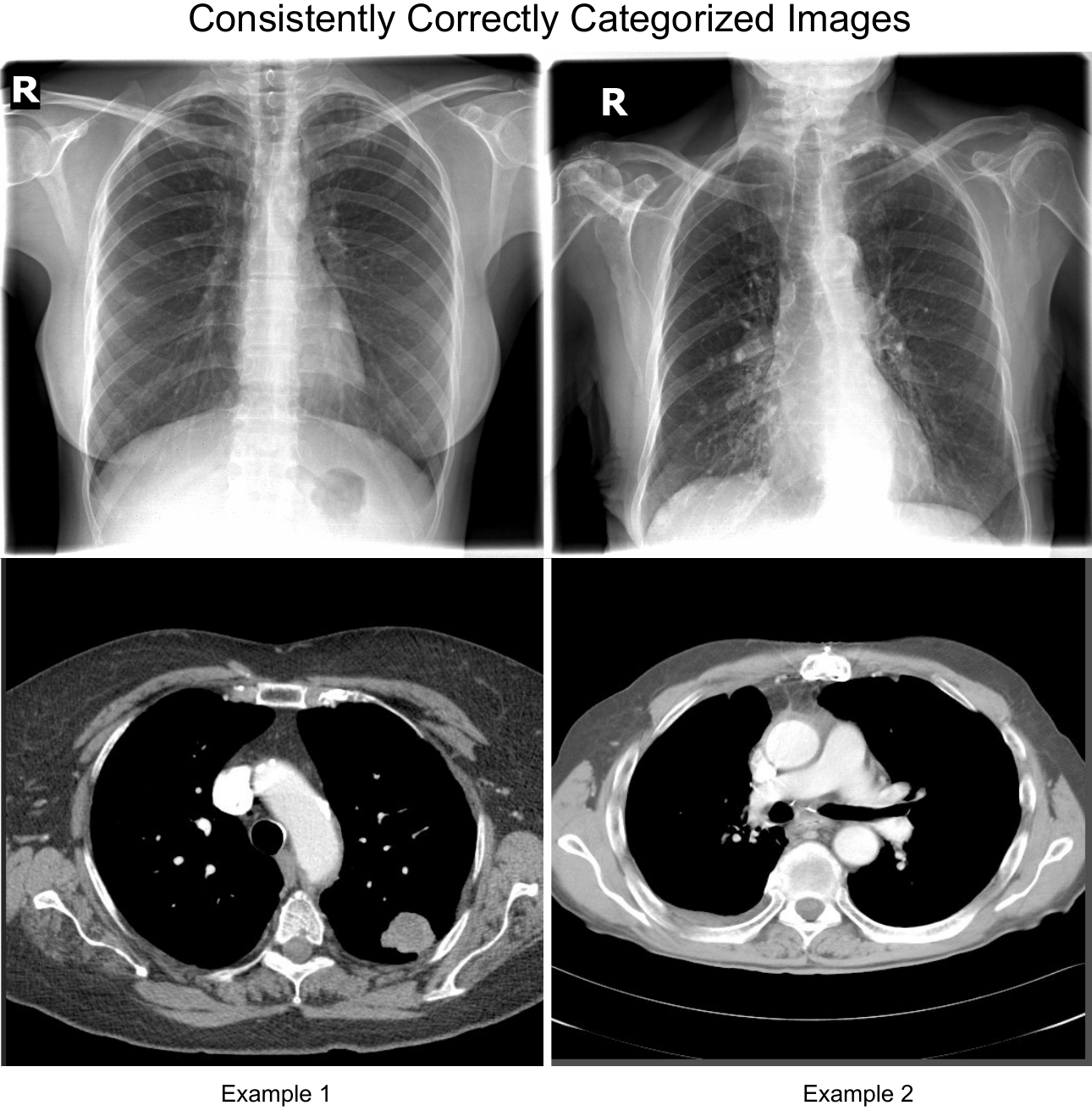}
    \includegraphics[width=1.0\linewidth]{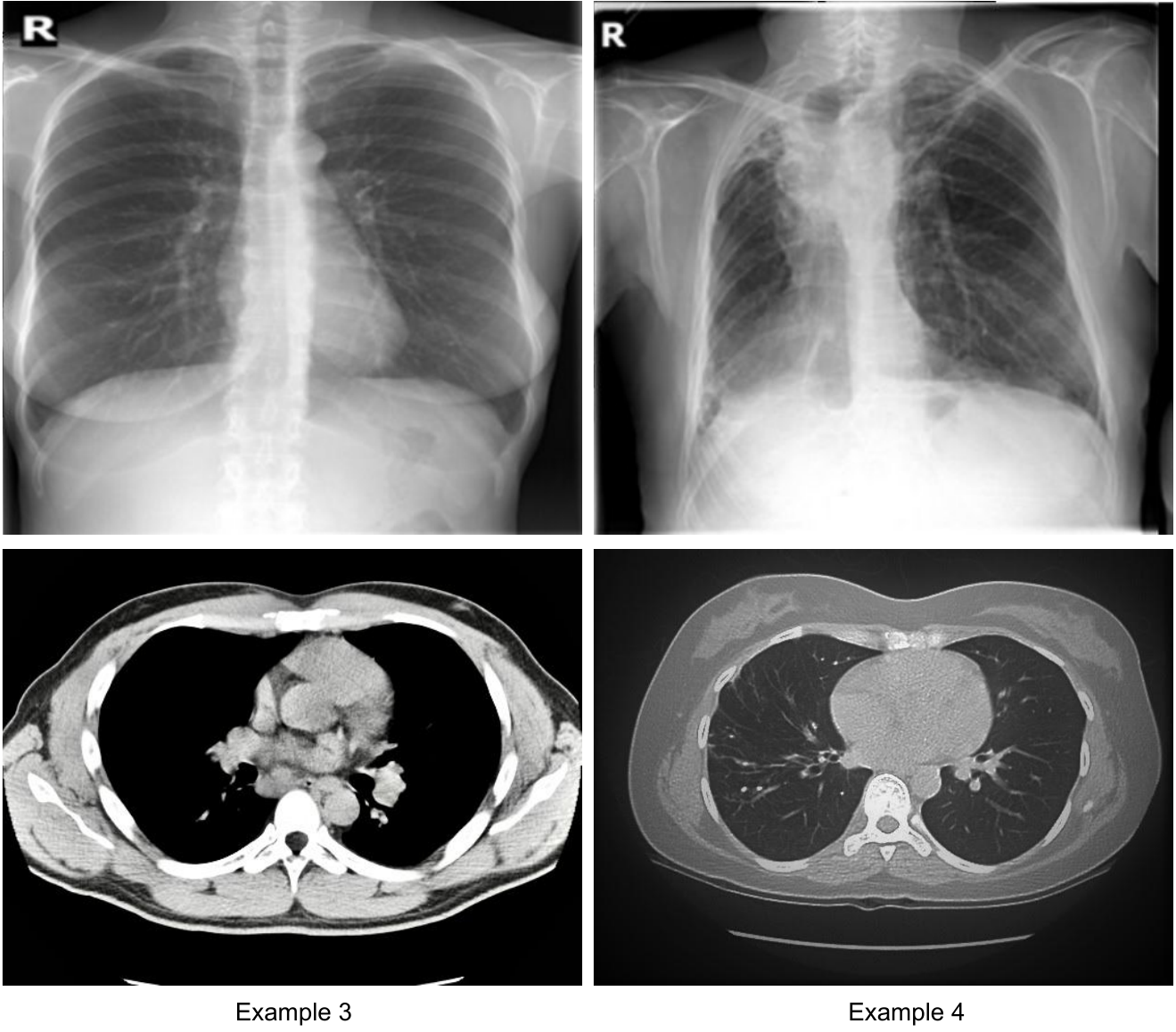}
    \caption{\emph{Sample examples of PA X-rays and corresponding generated CT scans that were most often correctly categorized across all models.} Patient positioning, noise, and absence of internally placed devices typically contribute to more accurate predictions.}
    \label{fig:examples_success}
\end{figure}

\begin{figure}[h]
    \centering
    \includegraphics[width=1.0\linewidth]{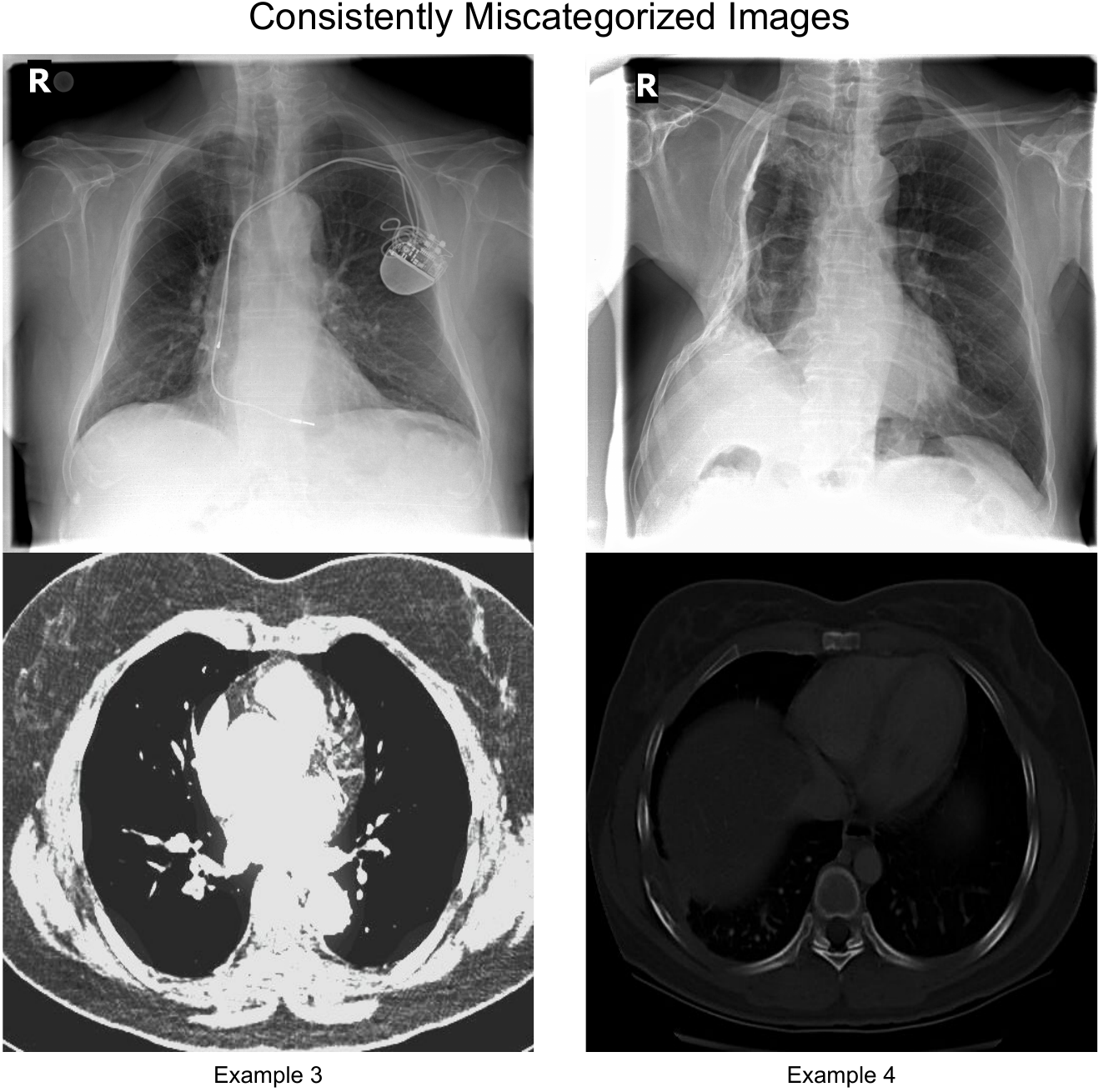}
    \includegraphics[width=1.0\linewidth]{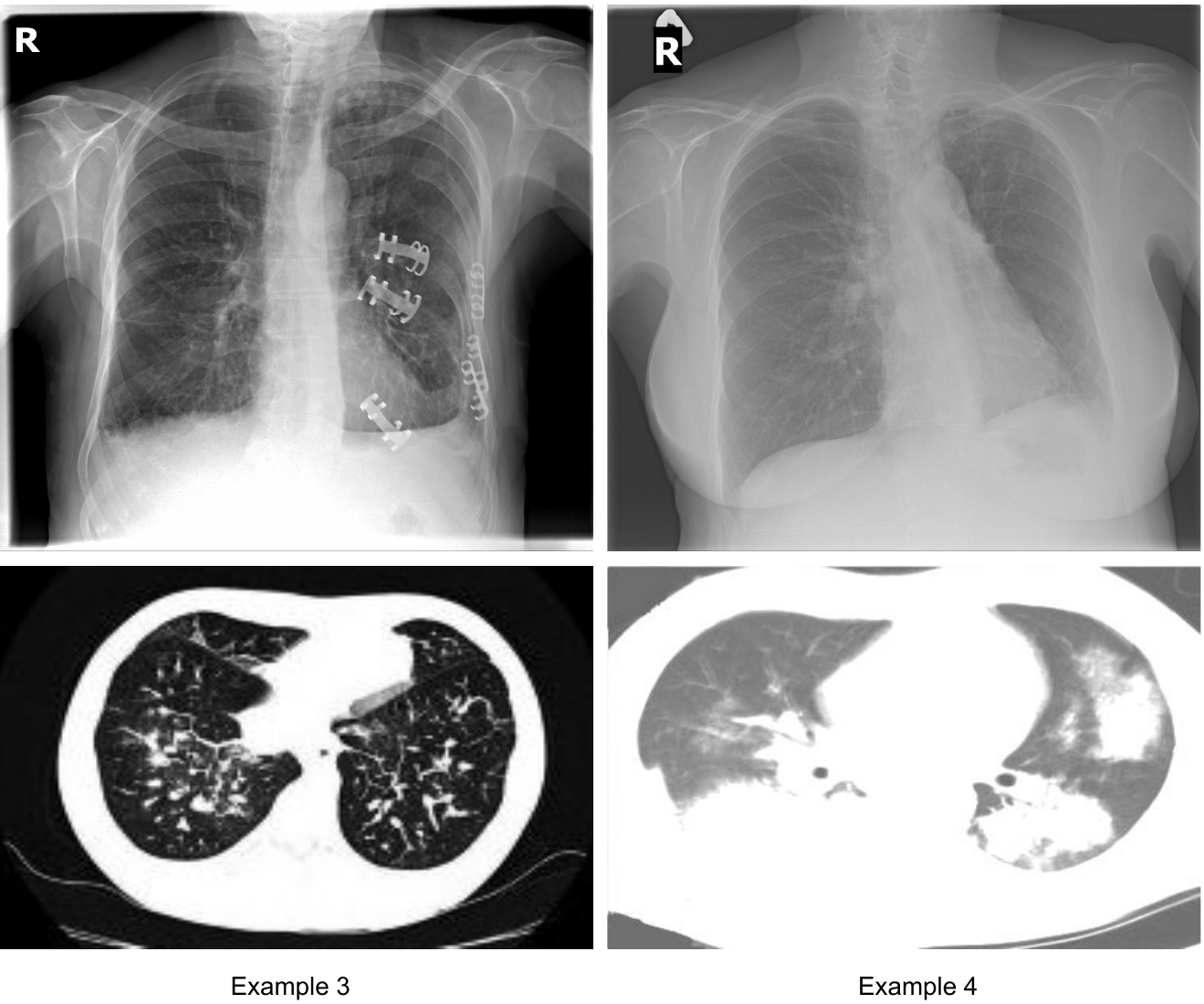}
    \caption{\emph{Sample examples of PA X-rays and corresponding generated CT scans that were most often incorrectly categorized across all models.} The first two images are apt representations of the x-rays that were often misrepresented. These x-rays often featured pacemakers and other devices in the image, or featured patients whose internal anatomy was significantly injured or anomalous. }
    \label{fig:examples_failure}
\end{figure}

\subsection{Qualitative Analysis}
\noindent Finally, we analyze the results qualitatively. In Figure~\ref{fig:examples_success}, we show sample examples of PA X-ray images and corresponding generated CT scans that are consistently correctly categorized. We find that data points that most consistently get categorized correctly represent patients with clear X-rays with little noise and no internal devices. In these cases, the CT generator also produces realistic CT images.

In contrast, we show some sample failure examples in Figure~\ref{fig:examples_failure}. Our system has difficulty with data points from patients whose X-rays show internal devices, such as pacemakers. Changes to internal anatomy, for instance, from injury, also presents a particular difficulty for both our approach and for the CT scan generator. For example, the CT generator in Example 4 produces a low-contrast CT that is missing details and hard to interpret. This observation highlights a known challenge of deploying a model trained on synthetic data to real world settings~\cite{ying2019x2ct}. 

\section{Conclusion and Future Work}
\noindent We present an automatic tuberculosis TB identification and classification approach that leverages synthetic data. By learning a computational model that predicts CT scans from X-rays, we improve disease classification in settings when no CT images are available. Experiments on a public benchmark demonstrate that our approach significantly improves accuracy of TB identification and classification. We also provide synthetic CT images for clinicians to be able to glean more insight than from Chest X-rays alone, who may not have access to this imaging. This approach opens the possibility of allowing regular follow up during the course of illness with synthetic chest CT imaging. Furthermore, chest X-rays could be compared with CT images more, as the cost of repeat Chest CT imaging is usually prohibitive.

There are a number of ways our work can be extended. First, the PadChest~\cite{padchest} dataset included many images that were not manually annotated by a radiologist or had a confirmed TB diagnosis. It would be important to test the proposed system on data where these labels are available and additionally performance could be compared across real and synthetic CT. We would also want to ensure a wide diversity in the images. Tuberculosis can come in many forms, including: miliary (scattered points throughout lung fields), effusions (fluid at base of lung, seen as white), and bronchogenic spread (with ill-defined fluffy air-space nodular opacities). This analysis was able to identify a wide range of presentations as TB. It will be important to ensure the dataset is balanced and includes rarer presentations, especially as some presentations may be more common in certain groups (particularly age groups or in persons with immunocompromise, such as with HIV). It will be important for the data to include a wider range of geographic and demographic origins, especially as presentation of TB may vary depending on diagnostic resources available and resultant time needed for diagnosis. In areas where TB care is obtained later in the course of illness, there may be more permanent damage and TB sequaelae will need lifelong follow-up for care. Ensuring a wide range of TB sequelae can help distinguish these cases from active disease and ensure more directed treatment. Quality of plain films can also vary and will be important to ensure films taken by a variety of devices work well. 

It would also be valuable to investigate how the technique can be applied for the study of other diseases. We hope that our explorations will encourage more research into the use of synthetic data to improve automatic disease diagnosis.

\section{Acknowledgements}
\noindent AL and EM were supported by the NYU Center for Data Science (CDS) Capital One Undergraduate Research Program (CURP) in partnership with the National Society of Black Physicists (NSBP). ES was supported by the Moore-Sloan Data Science Environment initiative (funded by the Alfred P. Sloan Foundation and the Gordon and Betty Moore Foundation) through the NYU Center for Data Science. We thank anonymous reviewers for their feedback.

{\small
\bibliographystyle{ieee_fullname}
\bibliography{egbib}
}

\end{document}